\documentstyle[12pt]{article}
\begin{document}
\title{Self--similarity in static axially symmetric relativistic fluids}
\author{L. Herrera$^{1}$\thanks{e-mail: lherrera@usal.es},
 and  A. Di Prisco$^{2}$\thanks{e-mail: alicia.diprisco@ciens.ucv.ve},\\
\small{$^1$ Instituto Universitario de Fisica Fundamental y Matematicas}\\
\small{Universidad de Salamanca, Salamanca, Spain}\\
\small{$^2$Escuela de F\'{\i}sica, Facultad de Ciencias,} \\
\small{Universidad Central de Venezuela, Caracas, Venezuela.}\\
\\
}
\maketitle

\vspace{-0.5cm}
\begin{abstract}
We carry on a general study on axially symmetric, static fluids admitting a  conformal Killing vector (CKV). The physical relevance of this kind of symmetry is emphasized. Next, we investigate all possible consequences derived from the imposition of such a symmetry. Special attention is paid to the problem of symmetry inheritance. Several families of solutions endowed with a CKV are exhibited.
 \end{abstract}

\section{Introduction}
In Newtonian hydrodynamics, self--similar solutions are those described by means of physical quantities which are functions depending on dimensionless variables $x/l(t)$, where $x$ and $t$ are independent space and time variables and $l$ is a time dependent scale. This implies that the spatial distribution of the characteristics of motion remains similar to itself at all times \cite{BZ}. Thus, self--similarity is to be expected whenever the system under consideration possesses no characteristic lenght scale. 

From this last comment it should be clear that self--similarity plays an important role in the study of systems close to the critical point, where the correlation length becomes infinite. In this case, different phases of the fluid (e.g. liquid--vapor) may coexist, the phase boundaries vanish and density fluctuations occur at all length scales. This last fact is vividly  exhibited in the critical opalescence. 

Also, examples of self--similar fluids may be found in the study of strong explosions \cite{S, S1, T} and thermal waves \cite{ZK, B, ZR}. 

In general relativity, self--similar solutions are related to the existence of a homothetic Killing vector field (HKV), which imposes specific restrictions on the metric tensor. A related issue is the existence of a conformal Killing vector field (CKV), which generalizes the condition imposed by a HKV. 

If this geometric similarity is extended to the physical variables as well, then we say that the corresponding symmetry is ``inherited'' by the matter variables. We shall focus our attention, on this issue.

Since the pioneering work by Cahill and Taub \cite{CT}, there has been a wealth of work done on the problem of self--similarity in general relativity, with especial emphasis  on the ensuing consequences from the existence of HKV  or CKV, and possible solutions to the Einstein equations (see for example \cite{H1}--\cite{shee} and references therein). This kind of symmetry has also been investigated in the context of other theories of gravitation (see for example \cite{RS}--\cite{DRGR}, and references therein). More recently, particular attention has been  paid to the modelling of wormholes admitting a one parameter group of CKV (see \cite{BHL1}--\cite{SIF2}, and references therein).

 Since most of the previous works on self--similarity are restricted to the spherically symmetric case, we shall consider here systems with a lower  degree of symmetry. We shall rule out cylindrical symmetry on physical grounds, since it implies unbounded sources. Thus we are left with axial symmetry. 

The rationale supporting, and encouraging, the study of  static axially symmetric sources,  is based on the well known fact that, according to the Israel theorem \cite{israel}, the Schwarzschild solution is, among all the exterior Weyl solutions \cite{weyl1}-\cite{weyln}, the only one  possessing a regular horizon. Thus, for  very compact objects, a bifurcation appears between any finite perturbation of  the Schwarzschild spacetime and any (exact) Weyl  metric (see \cite{i2}-\cite{in} and references therein for a discussion on this point). Therefore, whenever we wish to  study the quasi-spherical space–time resulting from the
fluctuations off Schwarzschild, it would be better off to  describe such deviations 
from an exact solution to the Einstein equations (of the Weyl family, if we restrict
ourselves to vacuum static, axially symmetric solutions) continuously linked to the
Schwarzschild metric through one of its parameters, instead of considering a perturbation
of the Schwarzschild space–time.

It is our purpose in this work to analyze the self--similarity  in axially symmetric relativistic fluids. Our interest in this study is twofold: on the one hand we would like to find out all possible consequences (physical and geometrical) derived from the assumption that the spacetime under consideration admits a  CKV. On the other hand, we shall search for solutions, admitting a  CKV as a heuristic condition.

The analysis of axially symmetric static fluids, has been considered in the past, with particular  emphasis in the search of exact solutions to the Einstein equations that could serve as sources of the Weyl metrics (see  \cite{1}--\cite{HHM}).

In most of these last references the  line element has been assumed to satisfy the so called Weyl gauge. However,  as we know, the Weyl gauge is obtained from the condition  $G^\rho_{\rho}+G^z_{z}=0$ (where $G$ denotes the Einstein tensor). Therefore, such a condition can always be satisfied in the vacuum (static and axially symmetric)  case, but imposes severe conditions for  the interior spacetime, since it implies $T^\rho_{\rho}+T^z_{z}=0$.

Our study will be based on the general formalism developed for static axially symmetric fluids in \cite{B1},  without resorting to the Weyl gauge, and considering the most general matter content consistent with the symmetries of the problem.

The manuscript  is organized as follows: In the next section we shall review the basics of the general formalism described in \cite{B1}. Next, we  impose the existence of a CKV and obtain all the ensuing consequences derived from this symmetry, in the context of our fluid distribution. 

The special case of a  sub--family of CKV,  which are usually referred to as Special Conformal Killing Vector (SCKV) is considered in section IV. The inheritance problem for this specific case is analyzed in some detail in section V.  

Next, in section VI, for fluids admitting a SCKV, and  different restricted choices of the generating vector, we deduce the corresponding equations of state. In section VII we discuss about the consequences of admitting a CKV on the Weyl tensor.   In order to illustrate how new solutions admitting CKV  may be found, some exact solutions are exhibited in section VIII, which satisfy additional  specific restrictions. 
 A summary of the obtained results  are presented in section IX. Finally several appendices are  included containing useful formulae.
\section{The general description of static axially symmetric fluids: The metric, the source and relevant equations}
In what follows we shall briefly summarize the definitions and main equations for describing the structure of a static axially symmetric fluid. We shall heavily rely on \cite{B1}, therefore we shall omit many steps in the calculations, details of which the reader may  find in \cite{B1}.

\subsection{The metric}
We shall consider, static and axially symmetric sources. For such a system the most general line element in ``Weyl spherical coordinates'', reads:

\begin{equation}
ds^2=-A^2 dt^2 + B^2 \left(dr^2 +r^2d\theta^2\right)+D^2d\phi^2,
\label{1b}
\end{equation}
where $A, B, D$ are functions of $r$ and  $\theta$.

It should be stressed  that we are not assuming the Weyl gauge, implying that our line element is defined by three independent functions, unlike  the vacuum case where it is always possible to reduce the line element so that only two independent metric functions appear.

Let us next provide a full description of the source. 
\subsection{The source}

We shall consider the most general source, compatible with staticity and  axial symmetry.  Thus we may write for the 
 energy momentum tensor in the ``canonical'' form:
\begin{eqnarray}
{T}_{\alpha\beta}&=& (\mu+P) V_\alpha V_\beta+P g _{\alpha \beta} +\Pi_{\alpha \beta}.
\label{6bis}
\end{eqnarray}
The above is the canonical, algebraic decomposition of a second order symmetric tensor with respect to unit timelike vector, which has the standard physical meaning when $T_{\alpha \beta}$ is the energy-momentum tensor describing some energy distribution, and $V^\mu$ the four-velocity assigned by certain observer.

Then, it is clear that $\mu$ is the energy
density (the eigenvalue of $T_{\alpha\beta}$ for eigenvector $V^\alpha$),  whereas  $P$ is the isotropic pressure, and $\Pi_{\alpha \beta}$ is the anisotropic tensor. We are considering an Eckart frame  where fluid elements are at rest. It is worth noticing that the anisotropic tensor in this particular  case (static) is not related to shear viscosity, but to any of the many physical processes which may cause anisotropy in stellar matter (see \cite{report} for a discussion on this issue)

Thus, it is immediate to see that 

\begin{equation}
\mu = T_{\alpha \beta} V^\alpha V^\beta, 
\label{jc10}
\end{equation} 

\begin{equation}
P = \frac{1}{3} h^{\alpha \beta} T_{\alpha \beta},\quad   \Pi_{\alpha \beta} = h_\alpha^\mu h_\beta^\nu \left(T_{\mu\nu} - P h_{\mu\nu}\right), 
\label{jc11}  
\end{equation}

with $h_{\mu \nu}=g_{\mu\nu}+V_\nu V_\mu$.

Since, we choose the fluid to be comoving in our coordinates, then
\begin{equation}
V^\alpha =(\frac{1}{A}, 0, 0, 0); \quad  V_\alpha=(-A, 0, 0, 0).
\label{m1}
\end{equation}

Let us now  define a canonical  orthonormal tetrad (say  $e^{(a)}_\alpha$), by adding to the four--velocity vector $e^{(0)}_\alpha\equiv V_\alpha$, three spacelike unitary vectors 

\begin{equation}
e^{(1)}_\alpha\equiv K_{\alpha}=(0, B, 0, 0); \quad  e^{(2)}_\alpha\equiv L_{\alpha}=\left(0, 0, Br, 0\right),
\label{7}
\end{equation}
\begin{equation}
 e^{(3)}_\alpha\equiv S_{\alpha}=(0, 0, 0, D),
\label{3nb}
\end{equation}
with $a=0,\,1,\,2,\,3$ (latin indices labeling different vectors of the tetrad).

The  dual vector tetrad $e_{(a)}^\alpha$  is easily computed from the condition 
\begin{equation}
 \eta_{(a)(b)}= g_{\alpha\beta} e_{(a)}^\alpha e_{(b)}^\beta, \qquad e^\alpha_{(a)}e_\alpha^{(b)}=\delta^{(b)}_{(a)},
\end{equation}
where $\eta_{(a)(b)}$ denotes the Minkowski metric.

In the above, the tetrad vector $e_{(3)}^\alpha=(1/D)\delta^\alpha_\phi$ is parallel to
the Killing vector of the axial symmetry  (it is the unit tangent to the orbits of the
group of 1--dimensional rotations that defines axial symmetry). The other two
basis vectors $e_{(1)}^\alpha,\,e_{(2)}^\alpha$ define the two {\it unique}
directions that are orthogonal to the 4--velocity and to the Killing vector mentioned above.

In order to provide  physical significance to the components of the energy momentum tensor, it is instructive to  apply the Bondi approach \cite{Bo}.

Thus, following Bondi, let us introduce purely locally
Minkowski coordinates ($\tau, x, y, z$) (or equivalently,  consider a tetrad
field attached to such l.M.f.) by:
\begin{equation}
d\tau=Adt;\qquad dx=Bdr;\qquad dy=Br d\theta;\qquad dz=Dd\phi.
\label{2}
\end{equation}

Denoting by a hat  the components of the energy momentum tensor in such locally defined coordinate system, we have that the matter content is given by

\begin{equation}
\widehat{T}_{\alpha\beta}= \left(\begin{array}{cccc}\mu    &  0  &   0     &   0    \\0 &  P_{xx}    &   P_{xy}     &   0    \\0       &   P_{yx} & P_{yy}  &   0    \\0       &   0       &   0     &   P_{zz}\end{array} \right) \label{3},
\end{equation}
\\
where $\mu, P_{xy}, P_{xx}, P_{yy}, P_{zz}$ denote the energy density and different stresses, respectively, as measured by our locally defined Minkowskian observer. It is worth noticing that the off diagonal term $P_{xy}$ cannot be ruled out by the axial symmetry alone.

Also observe that  $P_{xy}= P_{yx} $ and, in general  $ P_{xx}  \neq  P_{yy}  \neq P_{zz}$.

Introducing
\begin{equation}
\hat V_\alpha=(-1,0,0,0);\quad \hat K_\alpha=(0,1,0,0);\quad  \hat L_\alpha=(0,0,1,0),
\label{4}
\end{equation}
we have
\begin{eqnarray}
\widehat{T}_{\alpha\beta}&=& (\mu+P_{zz})\hat V_\alpha \hat V_\beta+P_{zz} \eta _{\alpha \beta} +(P_{xx}-P_{zz})\hat K_\alpha \hat K_\beta\nonumber \\ &+& (P_{yy}-P_{zz})\hat L_\alpha \hat L_\beta +2P_{xy}\hat K_{(\alpha} \hat L_{\beta)}
\label{5},
\end{eqnarray}
where $\eta_{\alpha \beta}$ denotes the Minkowski metric.

Then transforming back to our coordinates, we obtain the components of the energy momentum tensor in terms of the physical variables as defined in the l.M.f.
\begin{eqnarray}
{T}_{\alpha\beta}&=& (\mu+P_{zz}) V_\alpha V_\beta+P_{zz} g _{\alpha \beta} +(P_{xx}-P_{zz}) K_\alpha  K_\beta\nonumber \\ &+& (P_{yy}-P_{zz}) L_\alpha L_\beta +2P_{xy} K_{(\alpha}  L_{\beta)}.
\label{6}
\end{eqnarray}

It would be useful to follow the notation in \cite{ADY}, and to express the anisotropic tensor   in the form 

\begin{eqnarray}
\Pi_{\alpha \beta}=\frac{1}{3}(2\Pi_I+\Pi_{II})(K_\alpha K_\beta-\frac{h_{\alpha
\beta}}{3})+\frac{1}{3}(2\Pi _{II}+\Pi_I)(L_\alpha L_\beta-\frac{h_{\alpha
\beta}}{3})+2\Pi _{KL}K_{(\alpha}L_{\beta)} \label{6bb},
\end{eqnarray}

with 

\begin{eqnarray}
 \Pi _{KL}=K^\alpha L^\beta T_{\alpha \beta},
 \quad  \label{7P}
\end{eqnarray}

\begin{equation}
\Pi_I=(2K^{\alpha} K^{\beta} -L^{\alpha} L^{\beta}-S^{\alpha} S^{\beta}) T_{\alpha \beta},
\label{2n}
\end{equation}
\begin{equation}
\Pi_{II}=(2L^{\alpha} L^{\beta} -S^{\alpha} S^{\beta}-K^{\alpha} K^{\beta}) T_{\alpha \beta}.
\label{2nbis}
\end{equation}

Comparing (\ref{6bis}) with (\ref{6}), we can very easily link the fluid variables appearing in (\ref{6bis}) with the physical variables measured by our l.M.f., thus we obtain:
\begin{eqnarray}
\Pi_{I}=2P_{xx}-P_{zz}-P_{yy}, 
\label{6bb}
\end{eqnarray}
\begin{equation}
\Pi_{II}=2P_{yy}-P_{xx}-P_{zz},
\label{6bbbn}
\end{equation}
and
\begin{equation}
P=\frac{P_{xx}+P_{yy}+P_{zz}}{3}, \quad \Pi _{KL}=P_{xy}.
\label{7P}
\end{equation}

\section{Conformal motions and the hydrodynamical variables}
We shall consider spacetimes whose line element is defined by (\ref{1b}), admitting a CKV, i.e. satisfying the equation
\begin{equation}
{\mathcal{L}}_\chi g_{\alpha \beta} =2\psi g_{\alpha \beta},
\label{1cmh}
\end{equation}
 where ${\mathcal{L}}_\chi$ denotes the Lie derivative with respect to the vector field ${\bf \chi}$, which unless specified otherwise, has the general form
\begin{equation}
{\bf \chi}=\alpha(t, r, \theta)\partial_t+\beta(t, r, \theta)\partial_r+\gamma(t, r, \theta)\partial_{\theta},
\label{2cmh}
\end{equation}
 and $\psi$ in principle is a function of $t, r, \theta$. The case $\psi=constant$ corresponds to a HKV.

To find out the constraints that our assumption imposes on the hydrodynamical variables, let us take the Lie derivative of the Einstein equations, 
\begin{equation}
{\mathcal{L}}_\chi (R_{\alpha \beta}-\frac{1}{2}g_{\alpha \beta}R)=-8\pi {\mathcal{L}}_\chi T_{\alpha \beta},
\label{ccm1}
\end{equation}
where
\begin{equation}
{\mathcal{L}}_\chi R_{\alpha\beta}=2\psi_{;\alpha\beta}+g_{\alpha\beta}\,g^{\gamma \delta}\psi_{;\gamma \delta},
\label{1ab}
\end{equation}

\begin{equation}
{\mathcal{L}}_\chi R=6\,g^{\alpha \beta}\psi_{;\alpha \beta}-2\psi R,
\label{1abc}
\end{equation}
producing
\begin{equation}
4\pi{\mathcal{L}}_\chi T_{\alpha \beta}=g_{\alpha \beta}\,g^{\gamma \delta}\psi_{;\gamma \delta}-\psi_{;\alpha \beta}.
\label{2a}
\end{equation}

All the scalar equations ensuing from (\ref{ccm1}), are obtained by projecting on all possible combinations of the tetrad vectors. These equations are displayed in the Appendix B.

We may further transform the above equations, by using  the fact that, for any four--vector $X^\alpha$ (timelike or spacelike), the following relationship holds:
\begin{equation}
{\mathcal{L}}_\chi X ^{\alpha} =-\psi X^{\alpha}+Y^\alpha,
\label{3cmhn}
\end{equation}
or
\begin{equation}
{\mathcal{L}}_\chi X _{\alpha} =\psi X_{\alpha}+ Y_\alpha,
\label{3cmh}
\end{equation}
if ${\bf \chi}$ is a CKV, and $Y^\alpha$ is orthogonal to $X^\alpha$ (see \cite{MMT, CTU}).

Thus we may write:
\begin{eqnarray}
{\mathcal{L}}_\chi V_{\alpha} &=&\psi V_{\alpha}+V_{\perp \alpha},\nonumber\\
{\mathcal{L}}_\chi K_{\alpha} &=&\psi K_{\alpha}+K_{\perp \alpha},\nonumber\\
{\mathcal{L}}_\chi L_{\alpha} &=&\psi L_{\alpha}+L_{\perp \alpha},
\label{LKLL}
\end{eqnarray}
where
\begin{eqnarray}
V^{\alpha}V_{\perp \alpha}=0;\qquad K^{\alpha}K_{\perp \alpha}=0;\qquad L^{\alpha}L_{\perp \alpha}=0.
\label{LKLLab}
\end{eqnarray}
Then from (\ref{6aa})--(\ref{8aa}), we obtain respectively:
\begin{eqnarray}
 {\mathcal{L}}_\chi P_{xx} +2\psi P_{xx}+2P_{xy}K^{\alpha} {\mathcal{L}}_\chi L_{\alpha}=
\nonumber\\
\frac{1}{4\pi}\left(g^{\alpha\beta}-K^{\alpha}K^{\beta}\right)\psi_{;\alpha\beta},\nonumber\\
\label{6aaa}
\end{eqnarray}

\begin{eqnarray}
 {\mathcal{L}}_\chi P_{xy} +2\psi P_{xy}
+\left(P_{xx}-P_{zz}\right)L^{\alpha} {\mathcal{L}}_\chi K_{\alpha}\nonumber\\
+\left(P_{yy}-P_{zz}\right)K^{\alpha} {\mathcal{L}}_\chi L_{\alpha}=
-\frac{K^{\alpha}L^{\beta}}{4\pi}\psi_{;\alpha\beta},\nonumber\\
\label{7aaa}
\end{eqnarray}

\begin{eqnarray}
{ \mathcal{L}}_\chi P_{yy} +2\psi P_{yy} +2P_{xy}L^{\alpha} {\mathcal{L}}_\chi K_{\alpha}=
\nonumber\\
\frac{1}{4\pi}\left(g^{\alpha\beta}-L^{\alpha}L^{\beta}\right)\psi_{;\alpha\beta}.\nonumber\\
\label{8aaa}
\end{eqnarray}
\section{Special conformal Killing vectors„  ($\psi_{;\alpha\beta}=0$)}
We shall now focus on a special sub--family of CKV, satisfying the condition: $\psi_{;\alpha\beta}=0$, which are usually referred to as Special Conformal Killing Vector (SCKV). This include of course the HKV  ($\psi =1$) and the KV  ($\psi=0$), cases.

Then assuming  $\psi_{;\alpha\beta}=0$, it follows at once from (\ref{3a})
\begin{equation}
 {\mathcal{L}}_\chi \mu + 2\psi\mu=0.
\label{3ps0}
\end{equation}

From (\ref{4aa})--(\ref{VSa})
\begin{eqnarray}
\left(P_{xx}-P_{zz}\right)V^{\alpha}K_{\perp \alpha}+P_{xy}V^{\alpha}L_{\perp \alpha}\nonumber\\
-\left(\mu+P_{zz}\right)K^\alpha V_{\perp \alpha}=0,
\label{4ps0p}
\end{eqnarray}

\begin{eqnarray}
\left(P_{yy}-P_{zz}\right)V^{\alpha} L_{\perp \alpha}+P_{xy}V^{\alpha}  K_{\perp \alpha}\nonumber\\
-\left(\mu+P_{zz}\right)L^\alpha  V_{\perp \alpha}=0,
\label{5aab}
\end{eqnarray}

\begin{equation}
\left(\mu+P_{zz}\right)S^\alpha V_{\perp \alpha}
=0.
\label{VSaap}
\end{equation}

From (\ref{6aaa})--(\ref{8aaa})
\begin{eqnarray}
 {\mathcal{L}}_\chi P_{xx} +2\psi P_{xx}+2P_{xy}K^{\alpha}L_{\perp\alpha}=0,
\label{6ps0p}
\end{eqnarray}

\begin{eqnarray}
{ \mathcal{L}}_\chi P_{xy} +2\psi P_{xy}
+\left(P_{xx}-P_{zz}\right)L^{\alpha} K_{\perp \alpha}\nonumber\\
+\left(P_{yy}-P_{zz}\right)K^{\alpha} L_{\perp \alpha}=0,
\label{7ps0p}
\end{eqnarray}

\begin{eqnarray}
 {\mathcal{L}}_\chi P_{yy} +2\psi P_{yy} +2P_{xy}L^{\alpha} K_{\perp \alpha}=0,
\label{8ps0}
\end{eqnarray}

and from (\ref{9aa})
\begin{eqnarray}
 {\mathcal{L}}_\chi P_{zz} +2\psi P_{zz} =0.
\label{9ps0}
\end{eqnarray}
In the above we have used (\ref{LKLL}) and (\ref{LKLLab}).

We shall next use these equations to tackle the problem of symmetry inheritance.
\section{Conditions for the symmetry inheritance}

We say that a  CKV is inherited by the matter distribution, if for any physical variable (say $M$) we have 
 ${\mathcal{L}}_\chi M +2\psi M =0$.
 We shall here investigate the conditions under which a SCKV is inherited, for the system under study. We shall consider two different situations: $P_{xy}=0$ and $P_{xy}\neq0$

\subsection{$P_{xy}=0$}
With $P_{xy}=0$ we obtain, from   (\ref{4ps0p})--(\ref{8ps0}), respectively:
\begin{eqnarray}
 {\mathcal{L}}_\chi P_{xx} +2\psi P_{xx}=0,
\label{6ps0ppy0}
\end{eqnarray}

\begin{eqnarray}
 {\mathcal{L}}_\chi P_{yy} +2\psi P_{yy} =0,
\label{8ps0ppy0}
\end{eqnarray}

\begin{equation}
\left(P_{xx}-P_{zz}\right)V^{\alpha}K_{\perp \alpha}-\left(\mu+P_{zz}\right)K^\alpha V_{\perp \alpha}=0,
\label{4ps0ppy0}
\end{equation}

\begin{equation}
\left(P_{yy}-P_{zz}\right)V^{\alpha}L_{\perp \alpha}-\left(\mu+P_{zz}\right)L^\alpha V_{\perp \alpha}=0,
\label{5ps0ppy0}
\end{equation}

\begin{eqnarray}
 \left(P_{xx}-P_{zz}\right)L^{\alpha} K_{\perp \alpha}\nonumber\\
+\left(P_{yy}-P_{zz}\right)K^{\alpha} L_{\perp \alpha}=0,
\label{7ps0ppy0}
\end{eqnarray}

\begin{equation}
S^\alpha V_{\perp \alpha}=0 \Longrightarrow V_{\perp \alpha} \quad{\rm lies\;  in \; the\;  plane\; KL}. 
\label{VSaap1}
\end{equation}

Therefore, taking into account (\ref{3ps0}) and (\ref{9ps0}), we see that in this particular case all the  non--vanishing fluid variables inherit the symmetry.

Furthermore, in order to  satisfy Eqs.(\ref{4ps0ppy0}), (\ref{5ps0ppy0}) and (\ref{7ps0ppy0}), we may consider the following subcases :
\begin{itemize}
\item Case {$P_{xx}=P_{yy}=P_{zz}$} (perfect, isotropic  fluid).
From Eqs. (\ref{4ps0ppy0}) and (\ref{5ps0ppy0}) we have 
\begin{eqnarray}
K^\alpha V_{\perp \alpha}=0, \nonumber\\
L^\alpha V_{\perp \alpha}=0, 
\label{siv}
\end{eqnarray}
which by virtue of  (\ref{VSaap1}) implies
\begin{eqnarray}
V_{\perp \alpha}=0 \Longrightarrow  {\mathcal{L}}_\chi V_{\alpha} =\psi V_{\alpha}.
\label{siv1}
\end{eqnarray}

\item Case {$P_{xx} \neq P_{zz}, \,P_{yy} \neq P_{zz}$}

\begin{eqnarray}
V^{\alpha}K_{\perp \alpha}=0, \, &\Rightarrow&  K_{\perp \alpha}\, {\rm in\, the\, plane\,} LS \nonumber\\
&\Rightarrow& K_{\perp \alpha}={ l}L_{\alpha}+{s_k}S_{\alpha},\nonumber\\
V^{\alpha}L_{\perp \alpha}=0, \,& \Rightarrow&  L_{\perp \alpha} \,{\rm in\, the\, plane\,} KS\nonumber\\
&\Rightarrow& L_{\perp \alpha}={ k}L_{\alpha}+{s_l}S_{\alpha},\nonumber\\
\label{LKLLa0}
\end{eqnarray}
then, to satisfy  (\ref{4ps0ppy0}) and (\ref{5ps0ppy0}), we reobtain Eqs.(\ref{siv}) and (\ref{siv1}), 
and Eq. (\ref{7ps0ppy0}) can be written as
\begin{eqnarray}
 \left(P_{xx}-P_{zz}\right)l
+\left(P_{yy}-P_{zz}\right)k=0,
\label{7ps0ppy0b}
\end{eqnarray}
\end{itemize}
which implies a specific constraint on the equation of state.

Alternatively , we may say that if we assume (\ref{siv1}), then (\ref{LKLLa0}) follows from   (\ref{4ps0ppy0}) and (\ref{5ps0ppy0}).
\subsection{$P_{xy}\neq0$}
If $P_{xy}\neq0$ then, for the matter variables to inherit the SCKV we must to assume:
\begin{eqnarray}
K^{\alpha}L_{\perp \alpha}&=&0,\nonumber\\
L^{\alpha}K_{\perp \alpha}&=&0.
\label{LKLLa}
\end{eqnarray}

Indeed, from the above conditions and (\ref{6ps0p}), (\ref{7ps0p}) and (\ref{8ps0}) we obtain
\begin{eqnarray}
 {\mathcal{L}}_\chi P_{xx} +2\psi P_{xx}=0,
\label{6ps0b}
\end{eqnarray}
\begin{eqnarray}
 {\mathcal{L}}_\chi P_{xy} +2\psi P_{xy}=0,
\label{7ps0b}
\end{eqnarray}
\begin{eqnarray}
 {\mathcal{L}}_\chi P_{yy} +2\psi P_{yy} =0.
\label{8ps0b}
\end{eqnarray}

In what follows we shall consider different assumptions about the form of the generator vector ${\bf \chi}$, to find out what kind of different contraints on the equations of state, appear in each case.

\section{Equations of state emerging from different types of  SCKV}
So far the generator vector $\bf \chi$  has been assumed to be of the most general form, given in (\ref{2cmh}). We shall next restrict the form of this vector, and we shall analyze the constraints imposed by such restrictions, on the equations of state of the fluid.

Before doing that, some useful expressions have to be found.

Thus, from the trace of the energy momentum tensor

\begin{equation}
T=g^{\alpha \beta}T_{\alpha \beta}=-\mu+P_{xx}+P_{yy}+P_{zz},
\label{Tr}
\end{equation}
we may easily find 
\begin{equation}
 {\mathcal{L}}_\chi T + 2\psi T=-2P_{xy}\left(K^{\alpha}  L_{\perp \alpha}+L^{\alpha} K_{\perp \alpha}\right),
\label{LTr}
\end{equation}
%with (\ref{LKLL})
%\begin{equation}
% \mathcal{L}_\chi T + 2\psi T=-2P_{xy}\left(K^{\alpha} L_{\perp \alpha}+L^{\alpha}  K_{\perp \alpha}\right)
%\label{LTra}
%\end{equation}
where  (\ref{3ps0}),(\ref{6ps0p}),(\ref{8ps0}) and (\ref{9ps0}) have been used.

Then, using (\ref{ccm1}),
\begin{equation}
{\mathcal{L}}_\chi g_{\alpha \beta} =2\psi g_{\alpha \beta}=\chi_{\alpha;\beta}+\chi_{\alpha;\beta},
\label{Lg}
\end{equation}
and the Bianchi identities, in the form
\begin{equation}
2 R^\alpha_{\; \beta;\alpha}=R_{,\beta},
\label{fBi}
\end{equation}
we have
\begin{equation}
\left(R^{\alpha}_{\; \beta} \chi^{\beta}\right)_{;\alpha}=-8\pi P_{xy}\left(K^{\alpha}  L_{\perp \alpha}+L^{\alpha}  K_{\perp \alpha}\right).
\label{Rcm}
\end{equation}

Therefore if the right hand side of (\ref{Rcm}) vanishes, the conservation law 
\begin{equation}
\left(R^{\alpha}_{\; \beta} \chi^{\beta}\right)_{;\alpha}=0,
\label{Rcm1}
\end{equation}
holds. 

It is worth noticing that the conditions (\ref{LKLLa}), which ensure the inheritance of the SCKV, are the same that lead to the conservation law above.

The expression (\ref{Rcm}) will be used below to deduce the equations of state related to different choices of  $\chi^\beta$.

\subsection{Case $\chi^\beta=\upsilon V^\beta$}
If we assume that ${\bf \chi}$ is parallel to the four velocity then, a direct calculation of $(R^{\alpha}_{\; \beta} \chi^{\beta})_{;\alpha}$  produces 

\begin{eqnarray}
\left(R^{\alpha}_{\; \beta} \chi^{\beta}\right)_{;\alpha}=8\pi \psi \left(\mu+P_{xx}+P_{yy}+P_{zz}\right)\nonumber\\
-8\pi P_{xy}\left(K^{\alpha}  L_{\perp \alpha}+L^{\alpha} K_{\perp \alpha}\right),
\label{vVa}
\end{eqnarray}
where (\ref{6}), (\ref{LKLL}), (\ref{Lg}) and the Einstein equations,  have been used.

Then,  it follows at once from  (\ref{Rcm}) and (\ref{vVa})
\begin{equation}
 \mu+P_{xx}+P_{yy}+P_{zz}=0 \Rightarrow \mu=-3P.
\label{stvV}
\end{equation}

\subsection{Case $\chi^\beta=\zeta S^\beta$}
In this case   we proceed exactly as in the previous one. We first calculate $(R^{\alpha}_{\; \beta} \chi^{\beta})_{;\alpha}$ for the vector $\bf \chi$ parallel to $\bf S$, obtaining
\begin{eqnarray}
\left(R^{\alpha}_{\; \beta} \chi^{\beta}\right)_{;\alpha}=8\pi \psi \left(-\mu+P_{xx}+P_{yy}-P_{zz}\right)\nonumber\\
-8\pi P_{xy}\left(K^{\alpha}  L_{\perp \alpha}+L^{\alpha}  K_{\perp \alpha}\right),
\label{vV}
\end{eqnarray}
which together with (\ref{Rcm}) produces
\begin{equation}
 -\mu+P_{xx}+P_{yy}-P_{zz}=0.
\label{stvV}
\end{equation}

\subsection{Case $\chi^\beta=\kappa K^\beta$}
In this case, the same routine produces 
\begin{eqnarray}
\left(R^{\alpha}_{\; \beta} \chi^{\beta}\right)_{;\alpha}=8\pi \psi \left(-\mu-P_{xx}+P_{yy}+P_{zz}\right)\nonumber\\
-8\pi P_{xy}\left(-K^{\alpha}  L_{\perp \alpha}+L^{\alpha}  K_{\perp \alpha}\right)-8\pi\left(\kappa P_{xy}  L^\alpha\right)_{;\alpha},\nonumber\\
\label{kK}
\end{eqnarray}
which using (\ref{Rcm}) becomes
\begin{eqnarray}
 \psi \left(-\mu-P_{xx}+P_{yy}+P_{zz}\right)=\left( \kappa P_{xy} L^\alpha\right)_{;\alpha}\nonumber\\
-2P_{xy}K^{\alpha} L_{\perp \alpha},
\label{stkK}
\end{eqnarray}
or
\begin{eqnarray}
 \psi \left(-\mu-P_{xx}+P_{yy}+P_{zz}\right)=\nonumber\\
\frac{1}{Br}\left[\left(\kappa P_{xy}\right)_{\theta}+\kappa P_{xy}\left(\frac{A_{\theta}}{A}+\frac{B_{\theta}}{B}+\frac{C_{\theta}}{C}\right)\right].
\label{stkKa}
\end{eqnarray}

\subsection{Case $\chi^\beta=\lambda L^\beta$}
Finally, if the vector $\bf \chi$ is parallel to $\bf L$, we have 
\begin{eqnarray}
\left(R^{\alpha}_{\; \beta} \chi^{\beta}\right)_{;\alpha}=8\pi \psi \left(-\mu+P_{xx}-P_{yy}+P_{zz}\right)\nonumber\\
-8\pi P_{xy}\left(K^{\alpha}  L_{\perp \alpha}-L^{\alpha}  K_\perp {\alpha}\right)-8\pi\left(\lambda P_{xy}  K^\alpha\right)_{;\alpha},\nonumber\\
\label{lL}
\end{eqnarray}
and  using (\ref{Rcm}) 
\begin{eqnarray}
 \psi \left(-\mu+P_{xx}-P_{yy}+P_{zz}\right)=\left(\lambda P_{xy} K^\alpha\right)_{;\alpha}\nonumber\\
-2P_{xy}L^{\alpha}  K_{\perp \alpha},
\label{stlL}
\end{eqnarray}
which can be written as
\begin{eqnarray}
 \psi \left(-\mu+P_{xx}-P_{yy}+P_{zz}\right)=\nonumber\\
\frac{1}{B}\left[\left(\lambda P_{xy}\right)^{\prime}+\lambda P_{xy}\left(\frac{A^{\prime}}{A}+\frac{\left(Br\right)^{\prime}}{Br}+\frac{D^{\prime}}{D}\right)\right].
\label{stlLa}
\end{eqnarray}

In the above, $\upsilon, \zeta, \kappa, \lambda$ are arbitrary functions of $t, r, \theta$, and  prime  and subscript $\theta$, denote derivatives with respect to $r$  and $\theta$ respectively.

\section{The Weyl tensor}

It could be instructive to find out the consequences of the symmetry under consideration (CKV), on the Weyl tensor. In fact, it is known that the integrabilty conditions of (\ref{1cmh}) require (see chapter 7 in \cite{KY})
\begin{equation}
{\mathcal{L}}_\chi C^\alpha_{\gamma \beta \rho}=0,
\label{5cmhn}
\end{equation}
or
\begin{equation}
{\mathcal{L}}_\chi C_{\alpha\gamma \beta \rho}=2\psi C_{\alpha\gamma \beta \rho},
\label{6cmhn}
\end{equation}
where $C_{\mu\alpha\nu\beta}$ denotes the Weyl tensor.

The components of the electric Weyl tensor can be obtained directly from its definition (the magnetic part vanishes identically), 
\begin{equation}
E_{\mu\nu}=C_{\mu\alpha\nu\beta}\,V^\alpha\, V^\beta.\label{8}
\end{equation}

In our case the electric part of the Weyl tensor,  is defined by three non--vanishing independent components. Thus, the electric part of the Weyl tensor may also be written as:

\begin{eqnarray}
E_{\alpha \beta}={\mathcal{E}}_1\left(K_\alpha L_\beta+L_\alpha K_\beta\right)
+{\mathcal{E}}_2\left(K_\alpha K_\beta-\frac{1}{3}h_{\alpha \beta}\right)+{\mathcal{E}}_3\left(L_\alpha L_\beta-\frac{1}{3}h_{\alpha \beta}\right), \label{13}
\end{eqnarray}

where explicit expressions for the three scalars ${\mathcal{E}}_1$, ${\mathcal{E}}_2$, ${\mathcal{E}}_3$ are given in the Appendix C.

If (\ref{1cmh}) and (\ref{3cmh}) are satisfied, we may write
\begin{equation}
{\mathcal{L}}_\chi h_{\alpha \beta}=2\psi h_{\alpha \beta}+V_{\perp \alpha}V_{\beta}+V_{\perp \beta}V_{\alpha}.
\label{4cmhn}
\end{equation}

Then (\ref{3cmh}), (\ref{8}) and (\ref{6cmhn}) produce
\begin{eqnarray}
{\mathcal{L}}_\chi E_{\alpha \beta}&=&C_{\alpha \mu \beta \nu}\left(V^{\nu}V^{\mu}_{\perp}+V^{\mu}V^{\nu}_{\perp}\right)\nonumber \\
&=&V_{\perp \delta}\left(V_{\alpha} E^{\delta}_{\beta}+V_{\beta} E^{\delta}_{\alpha}\right).
\label{7cmhn}
\end{eqnarray}

Using (\ref{13}) we find for the non--vanishing proyections of (\ref{7cmhn})

$\left(V^{\alpha}K^{\beta}\right):$
\begin{eqnarray}
{\mathcal{E}}_1 V^{\alpha} L_{\perp \alpha}+{\mathcal{E}}_2 V^{\alpha} K_{\perp \alpha}=-{\mathcal{E}}_1 L^{\alpha} V_{\perp \alpha}-{\mathcal{E}}_2 K^{\alpha} V_{\perp \alpha},\nonumber\\
\label{EVK}
\end{eqnarray}

$\left(V^{\alpha}L^{\beta}\right):$
\begin{eqnarray}
{\mathcal{E}}_1 V^{\alpha} K_{\perp \alpha}+{\mathcal{E}}_3 V^{\alpha} L_{\perp \alpha}=-{\mathcal{E}}_1 K^{\alpha} V_{\perp \alpha}-{\mathcal{E}}_3 L^{\alpha} V_{\perp \alpha},\nonumber\\
\label{EVL}
\end{eqnarray}

$\left(K^{\alpha}S^{\beta}\right):$
\begin{eqnarray}
{\mathcal{E}}_1 S^{\alpha} L_{\perp \alpha}+{\mathcal{E}}_2 S^{\alpha} K_{\perp \alpha}=0,
\label{EKS}
\end{eqnarray}

$\left(L^{\alpha}S^{\beta}\right):$
\begin{eqnarray}
{\mathcal{E}}_1 S^{\alpha} K_{\perp \alpha}+{\mathcal{E}}_2 S^{\alpha} L_{\perp \alpha}=0,
\label{ELS}
\end{eqnarray}

$\left(K^{\alpha}L^{\beta}\right):$
\begin{eqnarray}
{\mathcal{L}}_\chi {\mathcal{E}}_1+2\psi {\mathcal{E}}_1+{\mathcal{E}}_2 L^{\alpha} K_{\perp \alpha}+{\mathcal{E}}_3 K^{\alpha} L_{\perp \alpha}=0,
\label{EKL}
\end{eqnarray}
while the combination of the  $\left(K^{\alpha}K^{\beta}\right)$, $\left(L^{\alpha}L^{\beta}\right)$ and $\left(S^{\alpha}S^{\beta}\right)$ projections, produce:

\begin{eqnarray}
2{\mathcal{E}}_1 \left(K^{\alpha} L_{\perp \alpha}+L^{\alpha} K_{\perp \alpha}\right)=0,
\label{Ekls1}
\end{eqnarray}

\begin{eqnarray}
{\mathcal{L}}_\chi {\mathcal{E}}_2+2\psi {\mathcal{E}}_2+2{\mathcal{E}}_1 K^{\alpha} L_{\perp \alpha}=0,
\label{Ekls2}
\end{eqnarray}

\begin{eqnarray}
{\mathcal{L}}_\chi {\mathcal{E}}_3+2\psi {\mathcal{E}}_3+2{\mathcal{E}}_1 L^{\alpha} K_{\perp \alpha}=0.
\label{Ekls3}
\end{eqnarray}

Thus, the three scalars ${\mathcal{E}}_1, {\mathcal{E}}_2, {\mathcal{E}}_3$ inherit the symmetry under consideration if  $ L^{\alpha} K_{\perp \alpha}=K^{\alpha} L_{\perp \alpha}=0$. It is worth noticing that these are the same conditions  (\ref{LKLLa}), required for the inheritance of SCKV  by the matter variables.

\section{Exact Solutions}
We shall now illustrate how to find exact interior, static axially symmetric solutions to the Einstein equations, admitting a one parameter group of CKV. The number of possible solutions (regardless of its possible physical viability) is huge,  therefore we shall introduce further restrictions to specify some of them. We shall consider the four cases discussed in section VI,  for SCKV. Solutions  admitting a one parameter group of CKV, but not belonging to the SCKV subcase are also exhibited

\subsection{$\chi^{\mu}=\kappa K^{\mu}$}
Let us assume that $\chi^{\mu}$ is parallel to  $K^{\mu}$, then 
from
\begin{eqnarray}
{\mathcal{L}}_\chi g_{\alpha \beta}=\chi^{\mu}g_{\alpha \beta,\mu}+g_{\alpha \mu}\chi^{\mu}_{,\beta}+g_{\beta \mu}\chi^{\mu}_{,\alpha}=2\psi g_{\alpha \beta},
\label{dg}
\end{eqnarray}
and 
\begin{equation}
\chi^{\mu}=\kappa K^{\mu},
\label{chk}
\end{equation}
we obtain
\begin{eqnarray}
\psi&=&\frac{\kappa}{B}\frac{A^\prime}{A},\nonumber\\
\psi&=&\frac{\kappa^{\prime}}{B},\nonumber\\
\psi&=&\frac{\kappa}{B}\frac{(Br)^{\prime}}{Br},\nonumber\\
\psi&=&\frac{\kappa}{B}\frac{D^{\prime}}{D},
\label{1234}
\end{eqnarray}
and 
\begin{eqnarray}
\left(\frac{\kappa}{B}\right)^{.}&=&0,\nonumber\\
\left(\frac{\kappa}{B}\right)_{,\theta}&=&0,\nonumber\\
\left(\frac{\kappa}{B}\right)_{,\phi}&=&0.
\label{567}
\end{eqnarray}
where the overdot denotes derivative with respect to $t$.

From the first equation (\ref{567}) it follows at once that $\kappa=\kappa(r,\theta)$, implying because of (\ref{1234}) that $\psi$ is independent on time. This last restriction, of course, is due to the specific choice of $\chi^{\mu}$. In general $\psi$ may depend on $t$.

Using (\ref{1234}) we can write
\begin{eqnarray}
A&=&C_{1}(\theta)\kappa \label{as1}, \\
Br&=&C_{2}(\theta)\kappa \label{as2}, \\
D&=&C_{3}(\theta)\kappa \label{as3},
\end{eqnarray}
where, by virtue of (\ref{567}), it follows that $C_{2}(\theta)=C_2=constant$.

Then, from $\psi_{;\alpha \beta}=0$ with $\dot\psi=0$, we obtain
\begin{eqnarray}
\psi^{\prime}\frac{A^{\prime}}{A}+\frac{\psi_{\theta}}{r^2}\frac{A_{\theta}}{A}&=&0, \label{ap} \\
\psi^{\prime\prime}-\psi^{\prime}\frac{B^{\prime}}{B}+\frac{\psi_{\theta}}{r^2}\frac{B_{\theta}}{B}
&=&0, \label{bp}\\
\psi^\prime_{\theta}-\psi^\prime \frac{B_{\theta}}{B}-\psi_{\theta}\frac{(Br)^\prime}{Br}&=&0, \label{cp}\\
\frac{\psi_{\theta\theta}}{r^2}+\psi^{\prime}\frac{(Br)^\prime}{Br}-\frac{\psi_{\theta}}{r^2}\frac{B_{\theta}}{B}&=&0, \label{dp}\\
\psi^\prime D^\prime+\frac{D_\theta \psi_\theta}{r^2}=0.\label{D}
\end{eqnarray}
From (\ref{bp}) and (\ref{dp}) we obtain
\begin{equation}
\psi^{\prime\prime}+\frac{\psi_{\theta\theta}}{r^2}+\frac{\psi^\prime}{r}=0,
\label{bd1}
\end{equation}
which helps to provide explicit expressions for $\psi$.

Thus for example, let us consider the simplest solution to (\ref{bd1}):
\begin{equation}
\psi=1,
\label{hcas}
\end{equation}
this defines a HKV, which of course is a special case of SCKV.

Then all the equations (\ref{ap})--(\ref{D}), are identically satisfied, and we obtain from (\ref{1234})--(\ref{as3}).
\begin{eqnarray}
A&=&C_{A}(\theta)r^{2C_B} \label{as1b}, \\
B&=&\alpha r^{(2C_B-1)} \label{as2b}, \\
D&=&C_{D}(\theta)r^{2C_B} \label{as3b},\\
\kappa &=&C_\kappa r^{2C_B} \label{as4b},
\end{eqnarray}
where $\alpha\equiv 2C_B C_\kappa$, $C_B, C_\kappa$ are arbitrary constants, and  $C_A, C_D, $ are arbitrary functions of $\theta$.

Then, using the field equations (\ref{24})--(\ref{26}) we obtain for the physical variables:
\begin{eqnarray}
8\pi \mu&=&-\frac{1}{\alpha^2 r^{4C_B}} \left(4C_B^2+\frac{C_{D,\theta\theta}}{C_D}\right),
\label{mu}
\\
8\pi P_{xx}&=&\frac{1}{\alpha^2r^{4C_B}} \left(12C_B^2 + \frac{C_{A,\theta\theta}}{C_A}+\frac{C_{D,\theta\theta}}{C_D}+\frac{C_{A,\theta}}{C_A}\frac{C_{D,\theta}}{C_D}\right),
\label{px}
\\
8\pi P_{yy}&=&\frac{1}{\alpha^2r^{4C_B}} \left(4C_B^2+\frac{C_{A,\theta}}{C_A}\frac{C_{D,\theta}}{C_D}\right),
\label{py}
\\
8\pi P_{zz}&=&\frac{1}{\alpha^2 r^{4C_B}} \left(4C_B^2+\frac{C_{A,\theta\theta}}{C_A}\right).
\label{pxx}
\\
P_{xy}&=&0.\label{pxy}
\end{eqnarray}

As expected these matter variables satisfy the conditions  (\ref{3ps0}), (\ref{9ps0})--(\ref{8ps0ppy0}), i.e. they inherit the SCKV, and satisfy the equation of state (\ref{kK}).

More involved expressions of $\psi$ lead to different solutions, although not always beloging to the SCKV class, and therefore not satisfying the inheritance conditions.

Thus, for example, a partial solution to (\ref{bd1}) is:
\begin{equation}
\psi=\left(\frac{a}{r}+br\right)\sin{\theta},
\label{bd}
\end{equation}
where $a$ and $b$ are two arbitrary constants.

Introducing this solution into (\ref{ap})--(\ref{dp}) we have
\begin{eqnarray}
-\left(\frac{a}{r^2}-b\right)\sin{\theta}\frac{A^{\prime}}{A}+\frac{1}{r}\left(\frac{a}{r^2}+b\right)\cos{\theta}\frac{A_{\theta}}{A}&=&0,\nonumber\\ \label{aps} \\
\frac{2a}{r^3}\sin{\theta}+\left(\frac{a}{r^2}-b\right)\sin{\theta}\frac{B^{\prime}}{B}+\frac{1}{r}\left(\frac{a}{r^2}+b\right)\cos{\theta}\frac{B_{\theta}}{B}&=&0,\nonumber\\ \label{bps}\\
\frac{2a}{r^2}\cos{\theta}+\left(\frac{a}{r}+br\right)\cos{\theta}\frac{B^{\prime}}{B}-\left(\frac{a}{r^2}-b\right)\sin{\theta}\frac{B_{\theta}}{B}&=&0,\nonumber\\ \label{cps}\\
-\frac{2a}{r^3}\sin{\theta}-\left(\frac{a}{r^2}-b\right)\sin{\theta}\frac{B^{\prime}}{B}-\frac{1}{r}\left(\frac{a}{r^2}+b\right)\cos{\theta}\frac{B_{\theta}}{B}&=&0,\nonumber\\ \label{dps}
\end{eqnarray}
we can see inmediately that (\ref{bps}) and (\ref{dps}) are identical. 

In order to specify further the solution,  we assume $a=0$. Then  from (\ref{bps}) and (\ref{cps}) it follows at once that:
\begin{equation}
B_{\theta}=B^\prime=0 \Rightarrow B=C_B=constant.
\label{sola0}
\end{equation}
Then (\ref{567}) and (\ref{as2}) imply $\kappa=\kappa(r)=C_4r$, whereas (\ref{as1}), (\ref{as3}) and (\ref{aps}), produce:

\begin{equation}
A=C_Ar\cos \theta,\qquad  D=C_{D}(\theta)r,
\label{sola0a}
\end{equation}
where $C_4$ and $C_A$ are arbitrary constants and $C_{D}(\theta)$ is an arbitrary function of its argument.

Alternatively, if we assume $a\neq0$, $b=0$, then the corresponding solution is:
\begin{eqnarray}
A=\frac{C_{A} \cos{\theta}}{r}, \qquad B=\frac{C_{B}}{r^2}, \qquad D=\frac{C_{D}(\theta)}{r}.
\label{sol12}
\end{eqnarray}

For both cases we may write for the physical variables:

\begin{eqnarray}
8\pi \mu&=&-\Omega \left(1+\frac{C_{D,\theta\theta}}{C_D}\right),
\label{mu}
\\
8\pi P_{xx}&=&\Omega \left(2+\frac{C_{D,\theta\theta}}{C_D}-\tan{\theta}\frac{C_{D,\theta}}{C_D}\right),
\label{px}
\\
8\pi P_{yy}&=&\Omega \left(1-\tan{\theta}\frac{C_{D,\theta}}{C_D}\right),
\label{py}
\end{eqnarray}
\begin{equation}
P_{zz}=P_{xy}=0,
\label{pzxy}
\end{equation}
where 
\begin{eqnarray}
\Omega\equiv\frac{1}{r^2 C_B^2}, \qquad (a=0),
\label{oma}\\
\Omega\equiv\frac{r^2}{C_B^2}, \qquad (b=0).
\label{omb}
\end{eqnarray}

However,  the above solutions admit a CKV which is not a SCKV, since  (\ref{D}) has not been satisfied. Accordingly, these solutions do not belong to the SCKV case, which explains why the matter variables do not inherit the CKV. Indeed, if we impose the condition (\ref{D}), we obtain at once $\frac{C_{D,\theta \theta}}{C_D}=-1$, implying $\mu=0$.

In general, if we adopt for $\psi$ the form
\begin{equation}
\psi=f(r)\sin{\theta},
\label{nri1}
\end{equation}
where $f$ is an arbitrary function,  then the equation (\ref{bd1}) becomes

\begin{equation}
Z^\prime+Z^2+\frac{Z}{r}-\frac{1}{r}=0,
\label{nri2}
\end{equation}
where$f=e^{\int Z dr}$.

The above is a Riccati equation, which can be reduced to a Bernoulli equation if we know some partial solution of it.

Indeed, if $Z_1$ is a partial solution to (\ref{nri2}), then introducing the new variable $W$ as $Z=Z_1+W$, the above equation becomes
\begin{equation}
W^\prime+W^2+W\left(2Z_1+\frac{1}{r}\right)=0,
\label{nri3}
\end{equation}
which is a Bernoulli equation that  can be easily linearized by introducing the new variable $Y=\frac{1}{W}$, producing
\begin{equation}
Y^\prime-1-Y\left(2Z_1+\frac{1}{r}\right)=0.
\label{nri4}
\end{equation}
Thus, for any known partial solution to (\ref{nri2}) we may integrate (\ref{nri4}), and obtain an explicit form of $f$. However  we should insist, once again, that not any solution to (\ref{bd1}) decribes a SCKV, since, for this to be true, all  equations (\ref{ap})--(\ref{D}) have to be satisfied (as we have seen with the solution (\ref{mu})--(\ref{omb})).

No solutions admitting a SCKV exist for the cases: $\chi^\beta=\lambda L^\beta$, $\chi^\beta=\zeta S^\beta$, $\chi^\beta=\upsilon V^\beta$.
\\

\section{Conclusions}
We have deployed all the equations required for a comprehensive study on axially symmetric static fluids admitting a CKV. 

 Then we have focused on the inheritance problem, for the particular case of SCKV. Conditions for the inheritance  of this symmetry by the physical variables have been  found for different  cases.  It is worth emphasizing the important role played by the off diagonal component $P_{xy}$  in this issue. We recall that such off diagonal term also plays  a fundamental role in the exit of the fluid from the equilibrium state. More specifically, it  has been shown that the value of this term deviates from its value in equilibrium, at the earliest stages of evolution (see \cite{evol} for a discussion on this point).

Next, we have shown how different forms of the generator vector give rise to different equations of state. For these choices we have also found some exact solutions. The pathologies  exhibited by such solutions indicate that they are not suitable to describe the whole fluid distribution, but just part of it. Such pathologies should not discourage  the search of exact solutions admitting a CKV, since they have been found under very restrictive conditions. By imposing them we just wanted to illustrate the way to find solutions. 

In order to find  physically meaningful solutions, some of the mentioned  restrictions have to be relaxed or, different kind of restrictions have to be impossed e.g.:
\begin{itemize}
\item To choose $\psi$ not satisfying the SCKV condition, i.e $\psi_{;\alpha \beta}\neq 0$.
\item To choose the generator vector $\chi$ not to be collinear with any of the tetrad vectors.
\item Choose for $\psi$ a more general solution of (\ref{bd1}), instead of (\ref{bd}).
\item To assume that  besides the admittance of the CKV, the spacetime is conformally flat.
\end{itemize}
\section{Acknowledgments}

This  work  was partially supported by the Spanish  Ministerio de Ciencia e
Innovaci\'on under Research Projects No.  FIS2015-65140-P (MINECO/FEDER).

\appendix
\section{Einstein equations}
For the line element (\ref{1b}) and the energy momentum tensor given by (\ref{6}),  the Einstein equations  read:

\begin{eqnarray}
8\pi\mu=-\frac{1}{B^2}\left\{\frac{B^{\prime \prime}}{B}+\frac{D^{\prime \prime}}{D}+\frac{1}{r}(\frac{B^\prime}{B} +\frac{D^\prime}{D})-(\frac{B^\prime}{B})^2+\frac{1}{r^2}\left[\frac{B_{\theta \theta}}{B}+\frac{D_{\theta \theta}}{D}-(\frac{B_\theta}{B})^2\right] \right\}
\label{24}
\end{eqnarray}
\begin{eqnarray}
8\pi P_{xx}=\frac{1}{B^2}\left[\frac{A^\prime B^\prime}{AB}+ \frac{A^\prime D^\prime}{AD}+\frac{B^\prime D^\prime}{BD}+\frac{1}{r}(\frac{A^\prime}{A}+\frac{D^\prime}{D})+\frac{1}{r^2}(\frac{A_{\theta \theta}}{A}+\frac{D_{\theta \theta}}{D}-\frac{A_\theta B_\theta}{AB}+\frac{A_\theta D_\theta}{AD}-\frac{B_\theta D_\theta}{BD})\right],
\label{25}
\end{eqnarray}
\begin{eqnarray}
8\pi P_{yy}=\frac{1}{B^2}\left[\frac{A^{\prime \prime}}{A}+ \frac{D^{\prime \prime}}{D}-\frac{A^\prime B^\prime}{AB} +\frac{A^\prime D^\prime}{AD}-\frac{B^\prime D^\prime}{BD}+\frac{1}{r^2}(\frac{A_\theta B_\theta}{AB}+\frac{A_\theta D_\theta}{AD}+\frac{B_\theta D_\theta}{BD})\right]
\label{27}
\end{eqnarray}
\begin{eqnarray}
8\pi P_{zz}=\frac{1}{B^2}\left\{\frac{A^{\prime \prime}}{A}+ \frac{B^{\prime \prime}}{B}-(\frac{B^\prime}{B})^2+\frac{1}{r}(\frac{A^\prime}{A} +\frac{B^\prime}{B}) +\frac{1}{r^2}\left[\frac{A_{\theta \theta}}{A}+\frac{B_{\theta \theta}}{B}-(\frac{B_\theta}{B})^2\right]\right\},
\label{28}
\end{eqnarray}
\begin{eqnarray}
8\pi P_{xy}=\frac{1}{B^2}\left\{  \frac{1}{r}\left[-\frac{A^{\prime}_\theta}{A} -\frac{D^{\prime}_\theta}{D} +\frac{B_\theta}{B}\left(\frac{A^\prime}{A}+\frac{D^\prime}{D}\right)+\frac{B^\prime}{B}\frac{A_\theta}{A}+\frac{B^\prime}{B}\frac{D_\theta}{D}\right]+\frac{1}{r^2} (\frac{A_\theta}{A}+\frac{D_\theta}{D})\right\}.\label{26}
\end{eqnarray}

Also, the nonvanishing components of the conservation equations $T^{\alpha  \beta}_{;\beta}=0$ yield:
\begin{equation}
\dot \mu=0,
\label{21}
\end{equation}

and

\begin{eqnarray}
P^\prime_{xx}+\frac{A^{\prime}}{A}(\mu+P_{xx})+\frac{B^{\prime}}{B}(P_{xx}-P_{yy})+\frac{D^{\prime}}{D}(P_{xx}-P_{zz})\nonumber \\+\frac{1}{r}\left[\left(\frac{A_\theta}{A}+2\frac{B_\theta}{B}+\frac{D_\theta}{D}\right)P_{xy}+P_{xy,\theta}+P_{xx}-P_{yy}\right]=0,
\label{22}
\end{eqnarray}

\begin{eqnarray}
P_{yy,\theta}+\frac{A_{\theta}}{A}(\mu+P_{yy})+\frac{B_{\theta}}{B}(P_{yy}-P_{xx})+\frac{D_{\theta}}{D}(P_{yy}-P_{zz})\nonumber \\+r\left[\left(\frac{A^{\prime}}{A}+2\frac{B^{\prime}}{B}+\frac{D^{\prime}}{D}\right)P_{xy}+P^{\prime}_{xy}\right]+2P_{xy}=0.
\label{23}
\end{eqnarray}

Equation (\ref{21}) is a trivial consequence of the staticity, whereas (\ref{22}) and (\ref{23}) are the hydrostatic equilibrium equations.

\section{Projections}
Projecting (\ref{2a}) an all possible combinations of the tetrad vectors ${\bf V}, {\bf K}, {\bf L}, {\bf S}$ we obtain:

$\left(V^{\alpha}V^{\beta}\right):$
\begin{equation}
 {\mathcal{L}}_\chi \mu + 2\psi\mu= -\frac{1}{4\pi}\left(g^{\alpha\beta}+V^{\alpha}V^{\beta}\right)\psi_{;\alpha\beta},
\label{3a}
\end{equation}

$\left(V^{\alpha}K^{\beta}\right):$
\begin{eqnarray}
\frac{1}{3}\left(2\Pi_{I}+\Pi_{II}\right)V^{\alpha} {\mathcal{L}}_\chi K_{\alpha}+\Pi_{KL}V^{\alpha} {\mathcal{L}}_\chi L_{\alpha}+\nonumber\\
\left[\frac{1}{3}\left(\Pi_{I}+\Pi_{II}\right)-\left(\mu+P\right)\right]K^\alpha { \mathcal{L}}_\chi V_{\alpha}=-\frac{V^{\alpha}K^{\beta}}{4\pi}\psi_{;\alpha\beta},\nonumber\\
\label{4a}
\end{eqnarray}

$\left(V^{\alpha}L^{\beta}\right):$
\begin{eqnarray}
\frac{1}{3}\left(2\Pi_{II}+\Pi_{I}\right)V^{\alpha} {\mathcal{L}}_\chi L_{\alpha}+\Pi_{KL}V^{\alpha} {\mathcal{L}}_\chi K_{\alpha}+\nonumber\\
\left[\frac{1}{3}\left(\Pi_{I}+\Pi_{II}\right)-\left(\mu+P\right)\right]L^\alpha  {\mathcal{L}}_\chi V_{\alpha}
=-\frac{V^{\alpha}L^{\beta}}{4\pi}\psi_{;\alpha\beta},\nonumber\\
\label{5a}
\end{eqnarray}

$\left(V^{\alpha}S^{\beta}\right):$
\begin{eqnarray}
\left[\frac{1}{3}\left(\Pi_{I}+\Pi_{II}\right)-\left(\mu+P\right)\right]S^\alpha { \mathcal{L}}_\chi V_{\alpha}
=-\frac{V^{\alpha}S^{\beta}}{4\pi}\psi_{;\alpha\beta},\nonumber\\
\label{VS}
\end{eqnarray}

$\left(K^{\alpha}K^{\beta}\right):$
\begin{eqnarray}
 {\mathcal{L}}_\chi P +2\psi P +\frac{1}{3} {\mathcal{L}}_\chi{\Pi_{I}} - \frac{2 \psi}{3}\left(\Pi_{I}
+\Pi_{II}\right)\nonumber \\
+\frac{2}{3}\left(2\Pi_{I}+\Pi_{II}\right)K^{\alpha} {\mathcal{L}}_\chi K_{\alpha}+2\Pi_{KL}K^{\alpha} {\mathcal{L}}_\chi L_{\alpha}=
\nonumber\\
\frac{1}{4\pi}\left(g^{\alpha\beta}-K^{\alpha}K^{\beta}\right)\psi_{;\alpha\beta},\nonumber\\
\label{6a}
\end{eqnarray}

$\left(K^{\alpha}L^{\beta}\right):$
\begin{eqnarray}
{ \mathcal{L}}_\chi \Pi_{KL} +\Pi_{KL}\left(K^{\alpha} {\mathcal{L}}_\chi K_{\alpha}+L^{\alpha} {\mathcal{L}}_\chi L_{\alpha}\right)\nonumber \\
+\frac{1}{3}\left(2\Pi_{I}+\Pi_{II}\right)L^{\alpha} {\mathcal{L}}_\chi K_{\alpha}+\frac{1}{3}\left(2\Pi_{II}+\Pi_{I}\right)K^{\alpha} {\mathcal{L}}_\chi L_{\alpha}=
\nonumber\\
-\frac{K^{\alpha}L^{\beta}}{4\pi}\psi_{;\alpha\beta},\nonumber\\
\label{7a}
\end{eqnarray}

$\left(L^{\alpha}L^{\beta}\right):$
\begin{eqnarray}
 {\mathcal{L}}_\chi P +2\psi P +\frac{1}{3} {\mathcal{L}}_\chi{\Pi_{II}} - \frac{2 \psi}{3}\left(\Pi_{I}
+\Pi_{II}\right)\nonumber \\
+\frac{2}{3}\left(2\Pi_{II}+\Pi_{I}\right)L^{\alpha} {\mathcal{L}}_\chi L_{\alpha}+2\Pi_{KL}L^{\alpha} {\mathcal{L}}_\chi K_{\alpha}=
\nonumber\\
\frac{1}{4\pi}\left(g^{\alpha\beta}-L^{\alpha}L^{\beta}\right)\psi_{;\alpha\beta},\nonumber\\
\label{8a}
\end{eqnarray}

$\left(S^{\alpha}S^{\beta}\right):$
\begin{eqnarray}
 {\mathcal{L}}_\chi P +2\psi P -\frac{1}{3} {\mathcal{L}}_\chi{\left(\Pi_{I}+\Pi_{II}\right)} - \frac{2 \psi}{3}\left(\Pi_{I}
+\Pi_{II}\right)=\nonumber \\
\frac{1}{4\pi}\left(g^{\alpha\beta}-S^{\alpha}S^{\beta}\right)\psi_{;\alpha\beta}.\nonumber\\
\label{9a}
\end{eqnarray}

Or, alternatively, using the physical variables $P_{xx}, P_{zz}, P_{yy}, P_{xy}$, we obtain from (\ref{4a})--(\ref{9a}), respectively

\begin{eqnarray}
\left(P_{xx}-P_{zz}\right)V^{\alpha} {\mathcal{L}}_\chi K_{\alpha}+P_{xy}V^{\alpha} {\mathcal{L}}_\chi L_{\alpha
}\nonumber\\
-\left(\mu+P_{zz}\right)K^\alpha  {\mathcal{L}}_\chi V_{\alpha}=-\frac{V^{\alpha}K^{\beta}}{4\pi}\psi_{;\alpha\beta},
\label{4aa}
\end{eqnarray}

\begin{eqnarray}
\left(P_{yy}-P_{zz}\right)V^{\alpha} {\mathcal{L}}_\chi L_{\alpha}+P_{xy}V^{\alpha} {\mathcal{L}}_\chi K_{\alpha}\nonumber\\
-\left(\mu+P_{zz}\right)L^\alpha  {\mathcal{L}}_\chi V_{\alpha}=-\frac{V^{\alpha}L^{\beta}}{4\pi}\psi_{;\alpha\beta},
\label{5aa}
\end{eqnarray}

\begin{equation}
-\left(\mu+P_{zz}\right)S^\alpha  {\mathcal{L}}_\chi V_{\alpha}
=-\frac{V^{\alpha}S^{\beta}}{4\pi}\psi_{;\alpha\beta},
\label{VSa}
\end{equation}

\begin{eqnarray}
{\mathcal{L}}_\chi P_{xx} +2\psi P_{zz}+2\left(P_{xx}-P_{zz}\right)K^{\alpha} {\mathcal{L}}_\chi K_{\alpha}\nonumber\\
+2P_{xy}K^{\alpha} {\mathcal{L}}_\chi L_{\alpha}=
\frac{1}{4\pi}\left(g^{\alpha\beta}-K^{\alpha}K^{\beta}\right)\psi_{;\alpha\beta},\nonumber\\
\label{6aa}
\end{eqnarray}

\begin{eqnarray}
{\mathcal{L}}_\chi P_{xy} +P_{xy}\left(K^{\alpha} {\mathcal{L}}_\chi K_{\alpha}+L^{\alpha} {\mathcal{L}}_\chi L_{\alpha}\right)\nonumber \\
\left(P_{xx}-P_{zz}\right)L^{\alpha} {\mathcal{L}}_\chi K_{\alpha}+\left(P_{yy}-P_{zz}\right)K^{\alpha} {\mathcal{L}}_\chi L_{\alpha}=
\nonumber\\
-\frac{K^{\alpha}L^{\beta}}{4\pi}\psi_{;\alpha\beta},\nonumber\\
\label{7aa}
\end{eqnarray}

\begin{eqnarray}
 {\mathcal{L}}_\chi P_{yy} +2\psi P_{zz} +2\left(P_{yy}-P_{zz}\right)L^{\alpha} {\mathcal{L}}_\chi L_{\alpha}
\nonumber\\
+2P_{xy}L^{\alpha} {\mathcal{L}}_\chi K_{\alpha}=
\frac{1}{4\pi}\left(g^{\alpha\beta}-L^{\alpha}L^{\beta}\right)\psi_{;\alpha\beta},\nonumber\\
\label{8aa}
\end{eqnarray}

\begin{eqnarray}
{ \mathcal{L}}_\chi P_{zz} +2\psi P_{zz} =
\frac{1}{4\pi}\left(g^{\alpha\beta}-S^{\alpha}S^{\beta}\right)\psi_{;\alpha\beta}.\nonumber\\
\label{9aa}
\end{eqnarray}

\section{Expression for the components of the electric Weyl tensor} 
For the three scalars ${\mathcal{E}}_1, {\mathcal{E}}_2, {\mathcal{E}}_3$ we obtain:

\begin{eqnarray}
{\mathcal{E}}_1= \frac{1}{2B^2} \left[\frac{1}{r}(\frac{A^{\prime}_\theta}{A} -\frac{D^{\prime}_\theta}{D}-
\frac{B_\theta}{B}\frac{A^{\prime}}{A}+\frac{D^{\prime}}{D}\frac{B_{\theta}}{B}-\frac{B^\prime}{B}\frac{A_\theta}{A}+
\frac{D_\theta}{D}\frac{B^\prime}{B})+\frac{1}{r^2}(\frac{D_{\theta}}{D}-\frac{A_\theta}{A})\right],\label{15}
\end{eqnarray}

\begin{eqnarray}
{\mathcal{E}}_2 & = &-\frac{1}{2B^2}\left[-\frac{A^{\prime \prime}}{A}+\frac{B^{\prime \prime}}{B}+
\frac{A^\prime B^\prime}{AB}+\frac{A^\prime D^\prime}{AD}-(\frac{B^\prime}{B})^2-\frac{B^\prime D^\prime}{BD}+\frac{1}{r}(\frac{B^\prime}{B}
-\frac{D^\prime}{D})\right]\nonumber \\ &-&\frac{1}{2B^2r^2}\left[\frac{B_{\theta \theta}}{B} -
\frac{D_{\theta \theta}}{D} -\frac{A_{\theta}B_{\theta}}{AB} +\frac{A_\theta D_\theta}{AD} -
(\frac{B_\theta}{B})^2+\frac{B_\theta D_\theta}{BD}\right],\label{16}
\end{eqnarray}

\begin{eqnarray}
{\mathcal{E}}_3 & = &-\frac{1}{2B^2}\left[\frac{B^{\prime \prime}}{B}-\frac{D^{\prime \prime}}{D}-
\frac{A^\prime B^\prime}{AB}+\frac{A^\prime D^\prime}{AD}-(\frac{B^\prime}{B})^2+\frac{B^\prime D^\prime}{BD}+\frac{1}{r}(\frac{B^\prime}{B}
-\frac{A^\prime}{A})\right]\nonumber \\ &-&\frac{1}{2B^2r^2}\left[\frac{B_{\theta \theta}}{B} -\frac{A_{\theta \theta}}{A}
 +\frac{A_{\theta}B_{\theta}}{AB} +\frac{A_\theta D_\theta}{AD} -
(\frac{B_\theta}{B})^2-\frac{B_\theta D_\theta}{BD}\right].\label{17}
\end{eqnarray}

Or, using Einstein equations we may also write:

\begin{eqnarray}
{\mathcal{E}}_1 =\frac{E_{12}}{B^2r}=4\pi P_{xy}+\frac{1}{B^2 r}\left[\frac{A^{\prime}_\theta}{A}-
\frac{A^{\prime} B_\theta}{AB}-\frac{A_\theta}{A} (\frac{B^{\prime}}{B}+\frac{1}{r})\right],
\label{18}
\end{eqnarray}

\begin{eqnarray}
{\mathcal{E}}_2 &=&-\frac{2E_{33}}{D^2}-\frac{E_{22}}{B^2r^2}=
{4\pi} (\mu+2P_{xx}+P_{yy})
-\frac{A^\prime}{B^2A}\left(\frac{2D^{\prime}}{D}
+\frac{B^{\prime}}{B}+\frac{1}{r}\right)\nonumber \\
&+&\frac{A_\theta}{AB^2r^2} \left(\frac{B_\theta}{B}-\frac{2D_\theta}{D}\right)-\frac{1}{B^2r^2}\frac{A_{\theta \theta}}{A},
\label{19}
\end{eqnarray}

\begin{eqnarray}
{\mathcal{E}}_3 =-\frac{E_{33}}{D^2}+\frac{E_{22}}{B^2r^2}=4\pi (P_{yy}-P_{zz})
-\frac{A^\prime}{B^2A}\left(\frac{D^{\prime}}{D}
-\frac{B^{\prime}}{B}-\frac{1}{r}\right)\nonumber \\
-\frac{A_\theta}{AB^2r^2} \left(\frac{D_\theta}{D}+\frac{B_\theta}{B}\right)+
\frac{1}{B^2r^2}\frac{A_{\theta \theta}}{A}.
\label{20}
\end{eqnarray}

\end{document}